# Dynamics of liquid crystal on hexagonal lattice

*Muhammad Arslan Shehzad[1,2], Junsu Lee[3], Sang Hoon Park,[1,2] Imtisal Akhtar[1,2], Muhammad Farooq Khan[1,3], Sajjad Hussain[1,2], Jonghwa Eom[1,3], Jongwan Jung[1,2], Gunn Kim[1,3]\*, Chanyong Hwang[4], Yongho Seo[1,2]\**

[1]*Graphene Research Institute, Sejong University, Seoul* 05006, *Republic of Korea*
[2]*Faculty of Nanotechnology & Advanced Materials Engineering, Sejong University, Seoul* 05006, *Republic of Korea*
[3]*Department of Physics, Sejong University, Seoul* 05006, *South Korea*
[4]*Center for Nano metrology, Korea Research Institute of Standards and Science, Daejeon 305-340, Korea*

**Abstract:**

Nematic liquid crystal (LC) molecules adsorbed on two dimensional materials are aligned along the crystal directions of the hexagonal lattice. It was demonstrated that short electric pulses can reorient the aligned LC molecules in the preferred armchair direction of hexagonal boron nitride (h-BN). Several states with a variety of colors were obtained by changing the direction and strength of the electric pulses. The ab initio calculations based on density functional theory was carried out to determine the favorable adsorption configurations of the LC molecules on the h-BN surface. A non-volatile display, in which pixel resolution can be determined by grains of hexagonal surface, is proposed, which can offer a pathway towards dynamic high-quality pixels with low power consumption, and could define a new paradigm for all non-volatile display applications.

# Introduction:

The key aspect of liquid crystal (LC) based devices is the interaction of the LC molecules with a surface that provides a force for orientational alignment.[1] These molecules have preferred orientations which depend upon the surface morphology, chemical modifications, and the external electric field.[2, 3] The director alignment is a vital factor for controlling the features of displays, which on a smooth surface can be either planar or homeotropic configuration.[2, 4] LC molecules have a tendency to be aligned on rubbed polymer films,[5] photo-polymerized structures,[6] and patterned layers.[7] Most of these techniques for controlling the orientation utilize morphological surface anisotropy. The controllability of these treatments depends on the surface structure and molecular arrangement at the submicron scale. It is still challenging to realize molecular-level control of the nematic LC molecules.

As conventional nematic LC molecules have preferential orientation spontaneously,[8, 9] an electric field should be applied to reorient them along the electric field against its preferential surface anchoring. To develop LC based non-volatile display devices showing static images without power consumption, it is required to have multistable states corresponding to the different colors.[10] If the device is set to a desired image that keeps its preferential alignment state, no power will be consumed until directors are switched for displaying a different image.[11]

Another key aspect linked to controllability is the stability of LC aligned states. While conventional nematic LC-based devices need color filters to show the color, LC devices with multistable orientations could provide flexibility to represent various colors without filter.[8] Earlier, some efforts were made to obtain bistable symmetry using intentionally fabricated square domains[12], where square patterns with four-fold symmetry provided two-fold symmetry of LC orientation. Similarly, using hexagonal patterns with six-fold symmetry yielded tristable symmetry.[10] Apart from these morphological anisotropies, it is known that the LC has tendency to be aligned along the hexagonal lattice crystals in bulk form,[13, 14] and their two-dimensional (2D) forms are attracting attention as they are transparent and flexible. Grains and grain boundaries of the van der Waals materials such as graphene, hexagonal boron nitride (h-BN), and transition-metal dichalcogenides are visualized using the LC's birefringence property.[15, 16] In the previous reports,[12,13] it was proposed that LC molecules tend to arrange along crystalline orientations of the 2D material owing to adsorptive interaction between the LC molecule and the hexagonal lattice.[16, 17] This technique was utilized to determine defects in graphene grown using chemical vapor deposition (CVD) during growth and transfer processes.[18] It is well-known that LC molecules tend to self-align into a helical structure with a small twist dependent on applied external force, which changes the polarization of propagating light along the helix exhibiting Bragg's reflection.[19]

In this paper, we report experimental and computational study of the flexibility of the alignment of LC molecules on 2D materials and proposed an innovative method to realize multistable states on a single grain of a hexagonal material. Taking the advantage of the insulating behavior of h-BN and hexagonal compatibility of LC molecules, alignment of LC molecules was tuned as a function of small DC voltage pulses. It has been demonstrated that small external force can re-orient the whole LC-domain along armchair directions. It was confirmed that several states with a variety of colors can be achieved by varying the combination of the electric field and the intensity of applied voltage pulses. It is believed that this approach can revolutionize the development of non-volatile and flexible displays, and also improve the utility of 2D nano-materials.

## Results and Discussion:

**Liquid crystal alignment:**

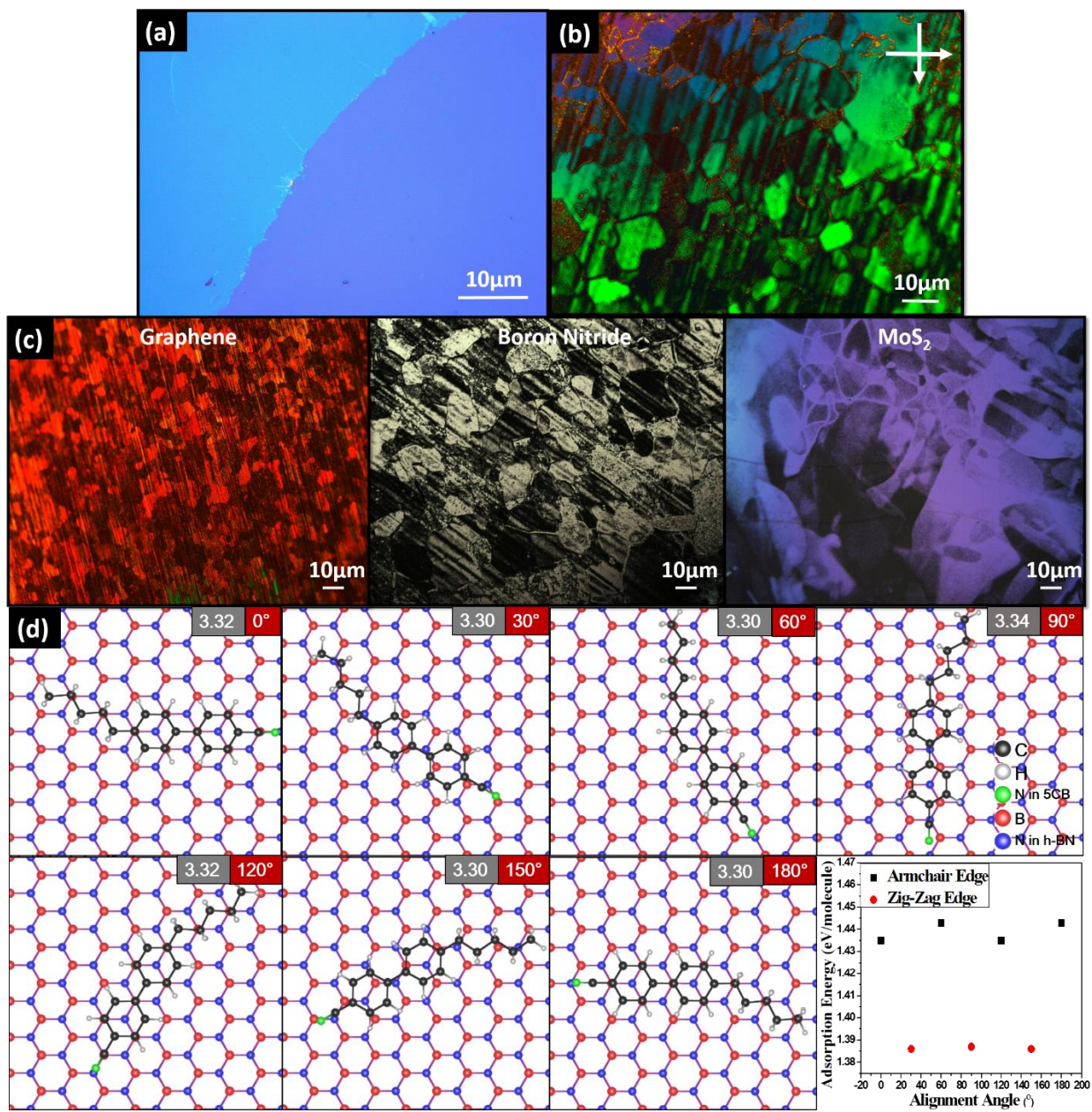

**Figure 1: LC alignment on 2D nano-materials:** CVD-grown 2D nano-materials were transferred to silicon. MoS$_2$ was directly grown on silicon; LC was molecularly aligned with atomic orientation of 2D nano-materials (a) Optical microscope image of transferred h-BN without LC. (b) POM with cross-polarizer image coated with LC. Noticeably, h-BN grains can be spotted and boundaries which were not visible without LC. (c) The same technique was applied to different 2D nano-materials (graphene and MoS$_2$), and domains were observed in all. LC molecules have different alignment angle depending on the orientation of grown grains which eventually show different colors. (d) Computational results on adsorption distance, angle, and adsorption energy for each configuration of the LC molecule on h-BN. Alignment along armchair edges have the highest adsorption energy, and zig-zag alignments also have similar values. Energy difference between zig-zag and armchair edges is only ~50 meV per molecule and was shown in bottom right of (d). The distances between 5CB molecule and h-BN sheet and adsorption energies are shown in legends.

CVD-grown graphene and h-BN were transferred onto a silicon substrate using a conventional wet chemical route, whereas $MoS_2$ was directly grown on the silicon surface (See Methods). LC (Sigma Aldrich-5CB) was spin coated to be 500 nm thick and heated above its isotropic temperature (60 ℃), and allowed to cool at a slow rate of 1 ℃/min. to minimize the effect of thermal flow.[20] Cooling leads to the alignment of molecules with a crystalline orientation of the hexagonal lattice structure. Figure 1(a) shows an optical microscope image of transferred h-BN without LC. A greenish color indicates the presence of h-BN and bluish color is ascribed to the oxide layer of the silicon. CVD-grown h-BN was characterized by atomic force microscopy (AFM) and Raman spectroscopy to estimate the quality and thickness of the transferred film (Figure S2). Figure 1(b) shows a polarized optical microscopy (POM) image of LC coated h-BN with a cross-polarizer. Prominently, grains and boundaries of h-BN were the evidence of molecular alignment of nematic LC. Growth of the h-BN was improved by tuning the growth time, and the LC alignment was confirmed either on partially or completely grown h-BN (Figure S3). The same technique was employed on graphene and molybdenum disulfide ($MoS_2$), and grains with clear boundaries were observed in all, as shown in Figure 1(c). The versatility of this technique confirms its utility to different 2D nano-materials with hexagonal symmetry.

Ab initio calculations based on density functional theory (DFT) have also been performed to determine the favorable adsorption configurations of a 4-cyano-4'-pentylbiphenyl (5CB) LC molecule on an h-BN sheet, using the Vienna ab initio simulation package (VASP)[21-24] (see methods). It was observed that h-BN has two different configurations with strong adsorption energy: armchair and zigzag directions with 120° rotation symmetry (Figure S4). The computational simulations reveal that AB stacking of two phenyl rings on the 2D structure has the highest adsorption energy, compared with other possible alignment configurations (Figure S5). As shown in Figure 1(d), considering AB stacking, we chose atomic model structures of the 5CB molecule physisorbed on the h-BN sheet, rotating the adsorbate through 180°. Although two phenyl rings in a 5CB molecule have a relative angle of ~30°, the rings in this study are practically parallel to each other so as to maximize the van der Waals interaction with the h-BN sheet. Four armchair directions (0, 60, 120, and 180°) and the three zigzag directions (30, 90, and 150°) was considered. Because of three-fold symmetry of h-BN, two structures with rotation angle difference of 120° correspond to an identical pair. All possible alignment states were observed and most stable states were depicted in Figure 1(d) (Figure S6). Consequently, the adsorption energies ($E_{ad}$) of the 5CB molecule, distance ($d$) between the 5CB molecule and h-BN sheet from the optimized atomic structures were derived. $E_{ad}$ is defined by

$$E_{ad} \equiv E_{5CB} + E_{hBN} - E_{5CB+hBN} \qquad \ldots (1)$$

Where, $E_{5CB}$, $E_{hBN}$ and $E_{5CB+hBN}$ are the total energies of free 5CB molecule, bare h-BN sheet, and a 5CB molecule absorbed on h-BN sheet, respectively. The distance, $d$, was defined by the difference between the average height of the 5CB molecule and h-BN sheet. Adsorption energies of the 5CB adsorbates aligned in the armchair directions were 50~60 meV/molecule higher than those of zigzag direction. This energy gap can be considered as an activation energy required for the 5CB molecule to rotate from one armchair direction to another. It was evident that there are three stable alignment states on a single grain, implying that LC molecules can be controlled into three possible directions by external sources. Earlier, we have suggested that the energy gaps were 25, 116 and 114 meV per molecule for graphene, MoSe$_2$ and WSe$_2$, respectively.[25] When two LC molecules were considered, including the interaction between them, a thermal excitation allows angle fluctuation, $\Delta\theta = \pm14.9°$, but cannot rotate the molecules from one armchair direction to another. (Figure S9)

To manifest the correlation between the LC director and crystalline orientation of the surface, the molecular interaction was investigated in depth previously. It was observed that LC molecules can align on the six preferential states of the hexagonal surface, which was further confirmed theoretically.[25] It shows that the anchoring direction was determined mainly by the axis of two phenyl rings parallel to the armchair direction of the hexagonal lattice. To check the stability of the alignment of the LC molecules, thermal cycling test, above the phase transition temperature (Figure S11, S12), was carried out. It was observed that the LC alignment directions were not completely the same as those before thermal treatment, indicating that the aligned direction of LC molecules can be changed by external conditions like temperature gradient, rather than determined by initial conditions like the geometry of the grains. The same alignment behavior was studied on bulk boron nitride flakes and interestingly, it was observed that LC can align on different orientation of single crystal h-BN. (Figure S13)

**LC-based display:**

To physically investigate the idea of multi alignment on h-BN, three electrodes forming a triangle were fabricated to apply different field strengths and directions. (Figure S14) After fabrication of the electrodes, h-BN was transferred without using polymer to avoid possible residue on the surface. Detailed processes were discussed in Supporting Information. (Figure. S15) LC was coated on the transferred film and the unidirectionally rubbed cover layer was placed on the LC (Figure. S16). Schematics of LC cell is shown in Figure 2(a), where the rubbed layer was placed on LC to align one end of LC layer (LC-glass interface) parallel to the polarizer direction. Figure 2(b) shows the degeneracy of alignment of LC molecules on the 2D surface along three preferential directions. Depending on the preferential alignment angles at the 2D surface, LC molecules form twisted states.

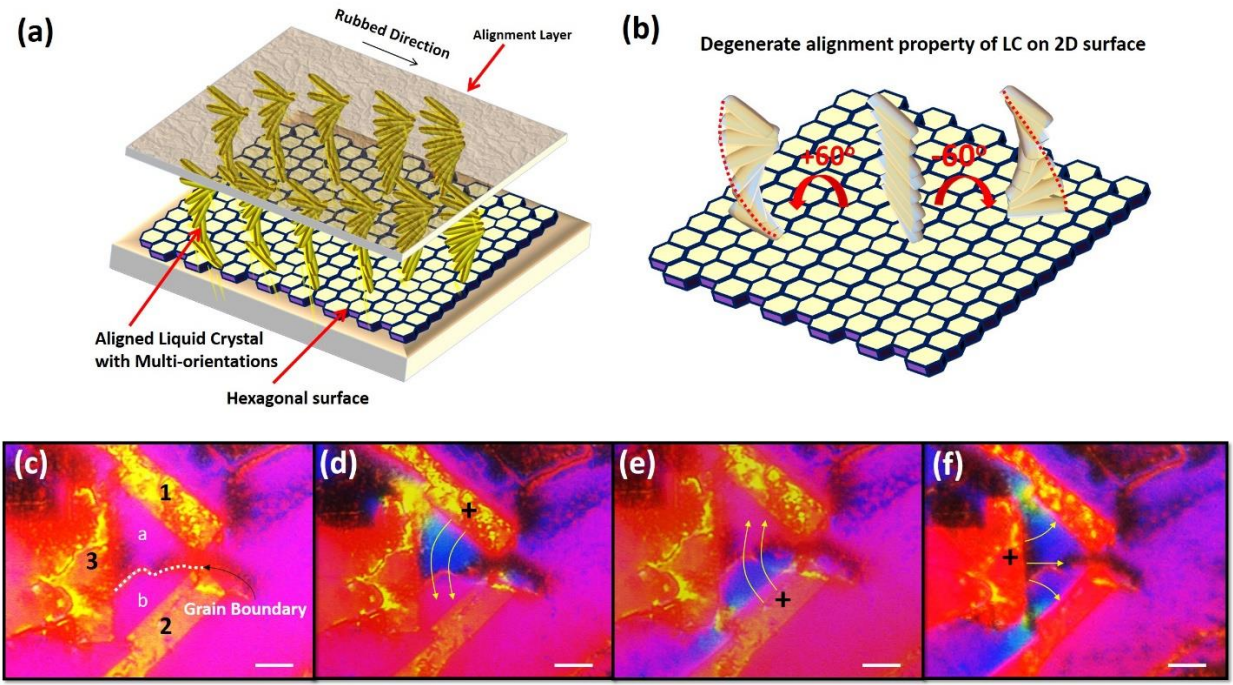

**Figure 2: Twisted states of LC on h-BN**. **(a)** LCs aligned on the transferred h-BN, and unidirectionally rubbed layer was placed on it. (b) The degeneracy of alignment of LC molecules on 2D surface causes different twisted states. (c) Two different grains of h-BN lay in this area with grain boundary (white line). (d) Short pulse of +10 V was applied on top electrode (with "+" marked) and other two electrodes were ground, and color of grain "a" was changed from red to blue. (e) When the second pulse was applied to other electrode (with "+" marked), the color of grain "b" became blue. (f) Once the third pulse was applied to third electrode, both grains ("a" and "b") were turned blue. The scale bars indicate 10 µm in all panels.

h-BN was transferred in such a way that two grains were located between the electrodes. As shown in Figure 2(c), grain boundary is marked with a white line between them. A pulse with +10V of 50 ms was applied to the electrode marked by "+" in Figure 2(d), while other two electrodes were grounded. Degenerate atomic compatibility of alignment on 2D surface assisted for LC to be twisted along one of the preferred orientations, which led to change red color of domain "a" to blue. In Figure 2(e), when field direction was changed by applying next pulse to other electrode, it also changed the color of domain "b" from red to blue. When the third pulse was applied at the third electrode covering the two grains ("a" and "b"), both of them were turned blue (See supporting video). As a result, LC domain on each grain can be controlled independently using external local field from neighboring electrode, and the repeatability was confirmed by consecutive experiments. Stability of each state was confirmed for an hour as shown in Figure S18. On the other hand, the gradation in color can be attributed to defects or impurities near the grain boundary, as well as the non-uniform thickness of the LC layer.

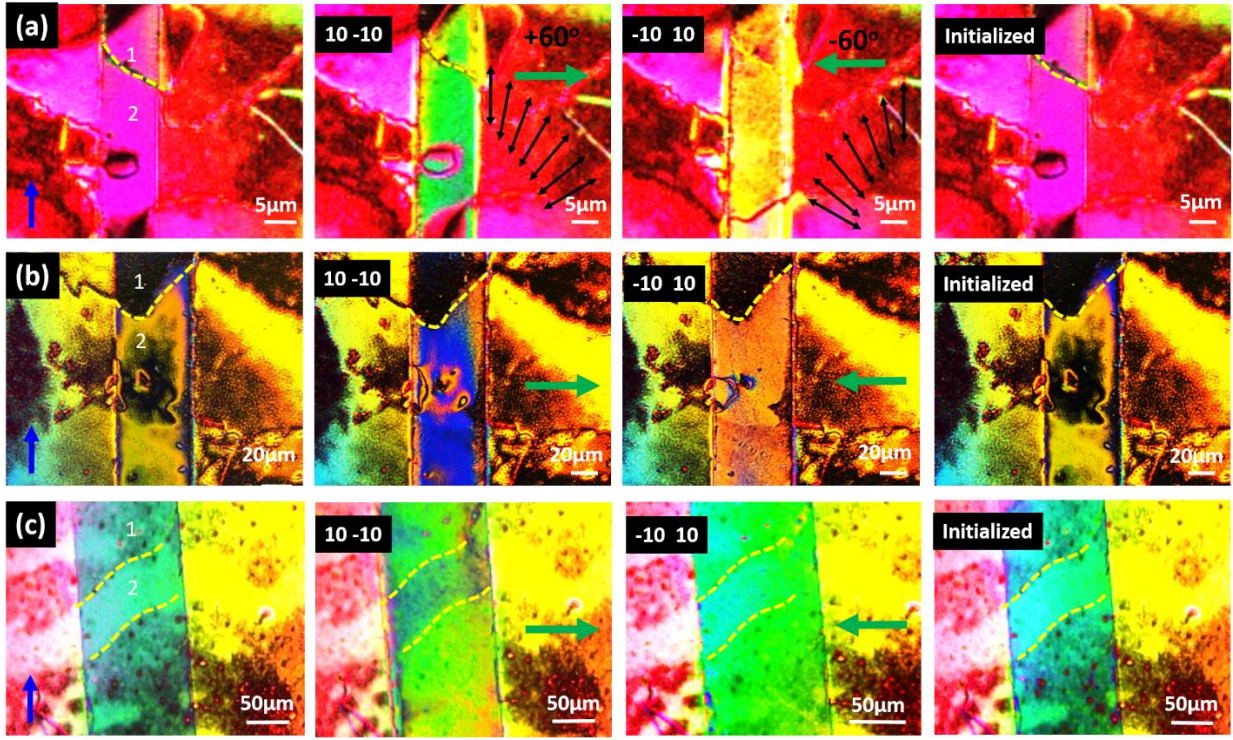

**Figure 3: Controlling three states with dynamic colors by pulse sequence on two electrodes:** Rubbed layer was placed on top with the rubbing direction parallel to the electrodes (blue arrows). Pulses with different field direction (marked with green arrows) were applied to reorient LC

directors as marked in the black arrows. Grain boundaries were marked with yellow guide lines. (a) When pulses of (10, -10 V) were applied on two electrodes with 10 µm gap, grains turned to green. When the polarity of the field was reversed (-10, 10 V) the grains changed their color from green to yellow. (b, c) The same experiments were performed on the samples with 50 and 100 µm gap electrodes and exhibited similar behaviors.

To confirm the versatility of this effect and to control the large area domain, two electrodes with different planar gaps were fabricated, as shown in Figure 3 (a-c). The LC layer thickness was adjusted to 1 µm with a rubbed layer, where the rubbing direction was indicated in blue arrows. DC pulses of 2.5 ms width with different polarities were applied. In Figure 3 (a), two grains were marked between electrodes with 10 µm spacing. When pulses of (10, -10 V) were applied on two electrodes, the grains 1 and 2 turned green. When polarity of the field was reversed (-10, 10 V) both the grains changed their color due to the opposite rotational twist of LC alignment. In order to revert back to the initial state, a vertical electric field (10 V) was applied using Si substrate, as shown in the last panel. Assuming most LC molecules were aligned nearly parallel to the rubbing direction initially, the LC alignment was tweeted by the pulses, as shown in the black arrows. When pulse was applied from the left to the right electrode, LC directors near the h-BN surface were rotated +60° towards the other preferential state, causing the LC layer to be twisted. Similarly, when field is applied from the right to the left, the LC directors were rotated -60° towards the third preferential state, causing twisting in the opposite direction. Interestingly, no color change was observed on the surface of the electrodes, which confirms the change was mainly dependent on the in-plane field between the electrodes.

Similar experiments were performed on the other samples having large grains, as shown in Figure 3(b-c). In Figure 3(b), two grains were marked between the electrodes with 50 µm gap, and the color of the 2nd grain was changed to blue (10 -10) and orange (-10 10) depending on the field direction. The grain 1 did not show any significant change, which can be explained by the original alignment direction. If the LC on the grain were initially oriented in either 0 or 180° to the external field, no torque is applied and no change in color occurs. Similarly, in Figure 3(c), the sample with large spacing (100 μm) also exhibited that the large domain of LC can be controlled and initialized by the same manner, even though there were some contaminations on the surface.

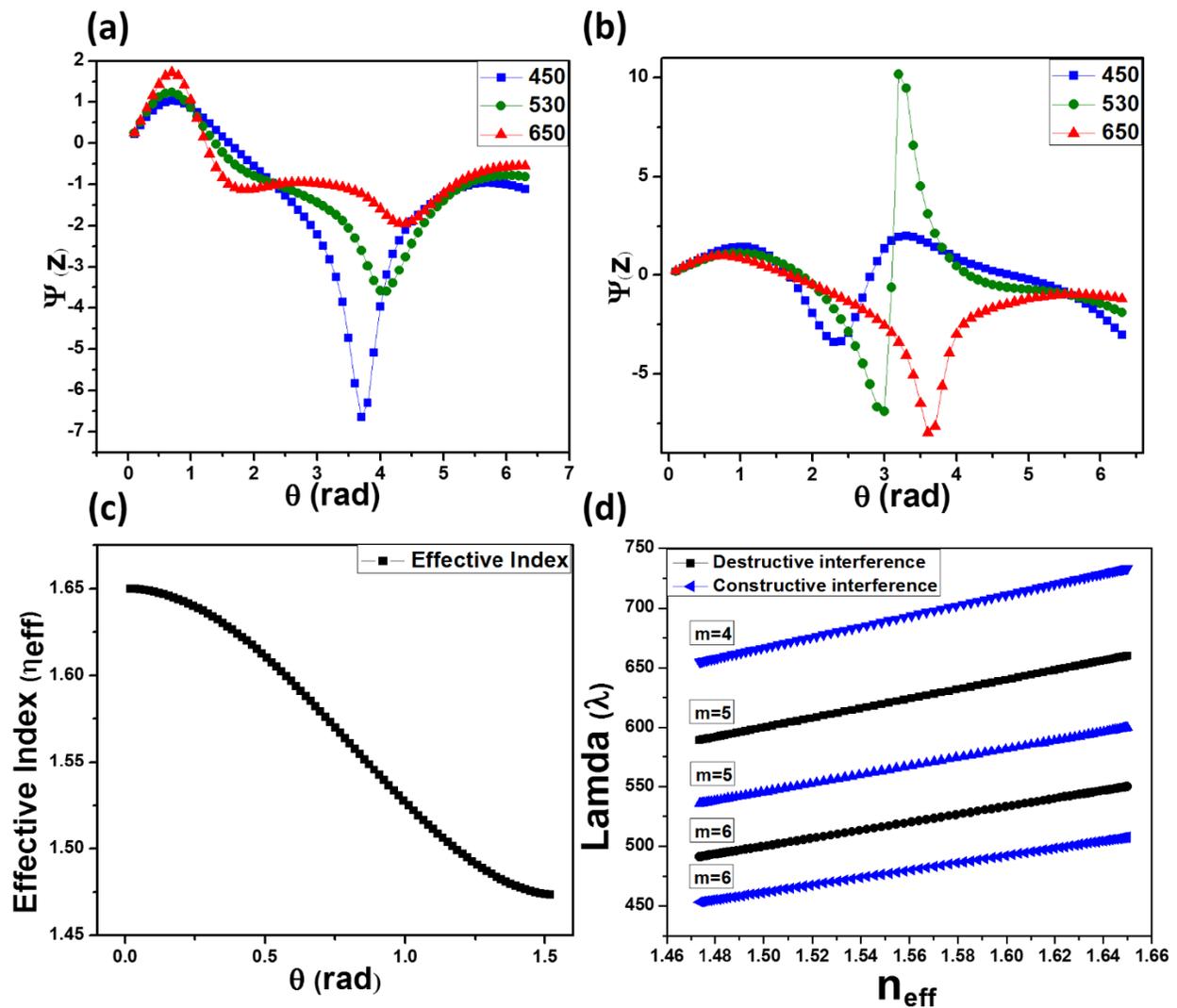

**Figure 4: Theoretical estimation of transmittance as function of twist angle:** (a-b) Relation between transmittance and twisted angles was plotted for various wavelengths ($\lambda = 450$, 530, and 600 nm), for the cells with widths (a) 2 μm and (b) 3 μm, respectively. The wavelength with strong intensity is varied depending on the twist angle. (c) The effective index of twisted LC due to birefringence was plotted as a function of twist angle. (d) The wavelengths causing constructive and destructive interferences of reflected beam from LC cell were plotted as functions of the effective index.

To understand the phenomena it is necessary to comprehend the dependency of transmittance on the twist angle and the wavelength. It is well-known that a periodic modulation of the helical structure of cholesteric liquid crystals results in a selective reflection band, which is referred to as a photonic crystal. The pitch of the rotation of conventional cholesteric LC is dependent on chiral dopant of LC.[26] In case of the nematic LC, the twist is induced by the boundary surfaces, and its twisting rate is much lower than that of cholesteric LC. Gooch and Tarry suggested a model to estimate the transmission intensity for the arbitrary twist angles between the two alignment layers.[27] Transmission of the light propagating through twisted LC cell is given as:

$$T = \frac{\sin^2[\theta\sqrt{1+u^2}]}{1+u^2} \quad \ldots \quad (2)$$

Where, $u = \frac{\pi d \Delta n}{\theta \lambda}$, $\theta$ is the twist angle due to different alignment layers, $d$ is the thickness of cell, $\Delta n = n_e - n_o$ is refractive index difference of liquid crystal, and $\lambda$ is the wavelength of light. In this model, the polarizer direction was rotated parallel to the alignment direction. In this study, however, the polarizer and analyzer were fixed perpendicular to each other, so Gooch-Tarry model should be modified. As LC molecules are aligned along armchair or zig-zag directions, $\theta$ is independently varied. From Azzam's model, when light propagates through the twisted LC, the ellipse of polarization, $\psi$, was described as:[28]

$$\psi(z) = V e^{2i\theta z/d} \quad \ldots \quad (3)$$

Here, $e^{2i\alpha z}$ represents periodic optical rotation, where, $z$ is propagating distance for the twisted LC cell located in x-y plane. The complex function, $V$, describes the variation of the polarization ellipse in the space-rotating coordinate system.[28] If the incident light is linearly polarized in the x-direction, the output polarization at $z = d$ is given as:

$$\psi(z) = \frac{\sqrt{1+u^2} - i(1+u)\tan(\theta\sqrt{(1+u^2)})}{\sqrt{1+u^2} + i(1-u)\tan(\theta\sqrt{(1+u^2)})} e^{2i\theta} \quad \ldots \quad (4)$$

In this experiment, the polarizer was fixed in x-axis, and the analyzer was in y-axis. Thus, the simplified general function of the transmittance can be obtained by taking imaginary parts (Details in Appendix):

$$\text{Im}(\psi) = \frac{\sqrt{a^2+c^2}}{\sqrt{a^2+b^2}} \sin\left(\tan^{-1}\frac{-c}{a} - \tan^{-1}\frac{b}{a} + 2\theta\right). \quad \ldots \quad (5)$$

Here, $a = \sqrt{1+u^2}$, $b = (1-u)\tan(a\,\theta)$, and $c = (1+u)\tan(a\,\theta)$. The function is plotted in Figure 4(a-b). The wavelength with maximum intensity is varied depending on the twist angle. Considering the reflected mode of cell, this effect might be further intensified. The polarized ellipse was also dependent on the thickness of the LC cell (Figure S20). As a result, the twist angle altered the wavelength with high intensity, and the LC domain exhibited different colors.

Additionally, the interference effect should be considered to explain the color change. A combination of constructive and destructive interference at different wavelengths causes different colors. Since the optical anisotropy of nematic LC causes birefringence and the effective refractive index is given as: [29]

$$n_{\text{eff}} = \frac{n_o n_e}{\sqrt{n_e^2 \sin^2\theta + n_o^2 \cos^2\theta}} \quad \ldots \quad (6)$$

Considering the ordinary refractive index, $n_o$, and extraordinary refraction index, $n_e$, for 5CB, the $n_{\text{eff}}$ was calculated (Figure 4(c)). It was evident that $n_{\text{eff}}$ was linearly dependent on twist angle.

Constructive ($2n_{eff}d = m\lambda$) or destructive ($2n_{eff}d = (m + 1/2)\lambda$) interference is expected depending on the $n_{eff}$ Figure 4(d) shows the change of wavelength depending on different refractive indices, satisfying Eq. (6), where d = 1 μm and m = 4, 5, 6. Twisting by a field changes the $n_{eff}$, which changes the wavelength of constructive or destructive interference, and different color was displayed. This interference effect influences the color change, which is complementary to the modified Gooch-Tarry model.

**Voltage-dependent tuning:**

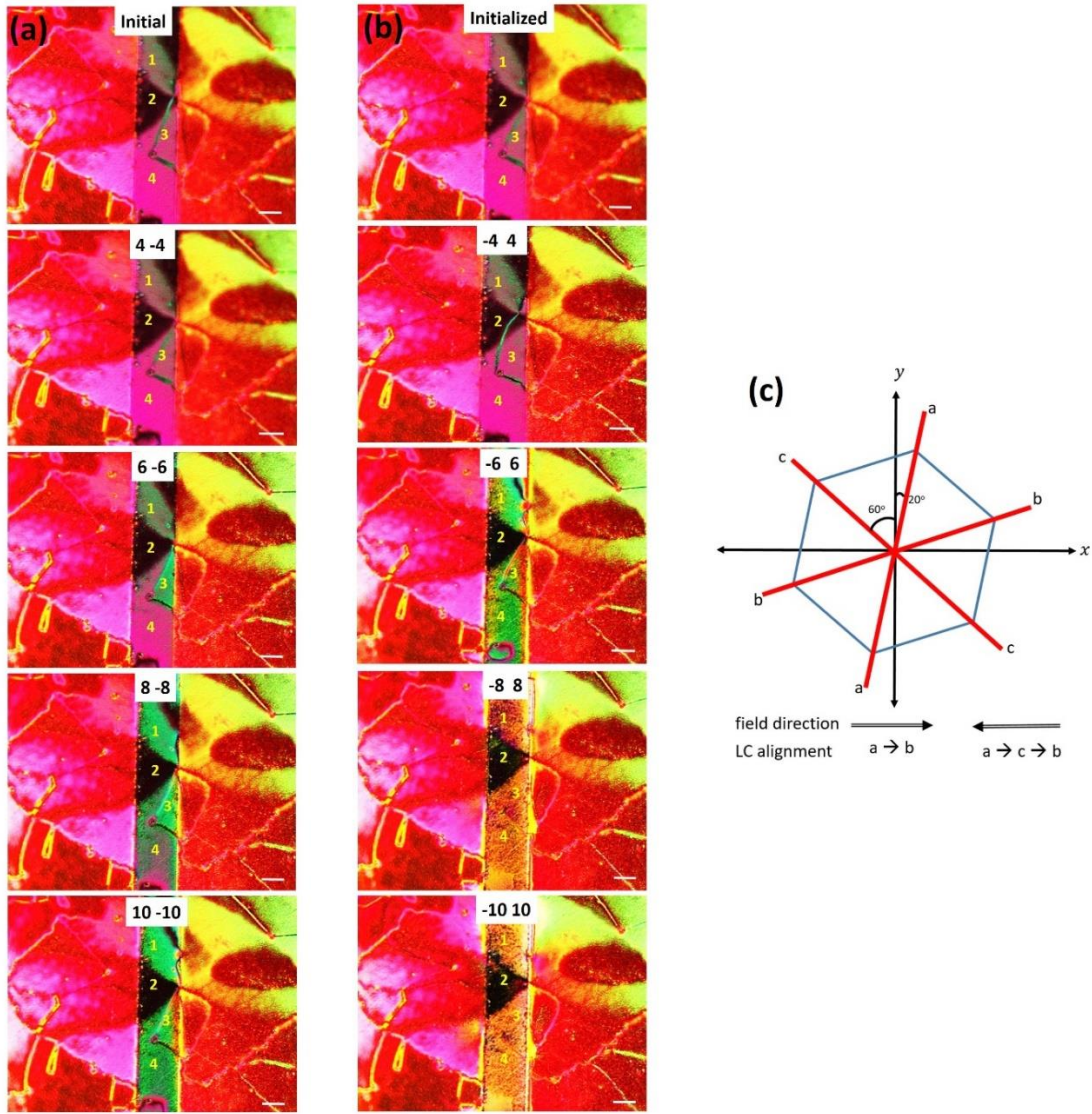

**Figure 5: Voltage dependent measurement:** To measure the threshold voltage, the pulse voltage was varied from -10 to 10 V. Four grains located in two parallel electrodes were numbered. (a) In case of field direction from left to right, grains 1 and 3 were converted to green around $\Delta V = 12$ V. By further increasing the field strength, grain 4 was converted to green. (b) When the field direction was changed, all three grains were converted to green at the same voltage. On further increase in the field strength, all three grains became orange. (c) To explain this behavior, a model with title crystal orientation of h-BN is suggested. The scale bars indicate 5 µm in all panels.

To estimate the threshold voltage $V_{th}$, pulses with different voltages up to $\pm 10$ V were applied transiently and POM images were taken as shown in Figure 5. In Figure 5(a), an electric field from left to right was applied between two parallel electrodes, and grains 1 and 3 were converted to green at around ΔV=12 V, which is approximately corresponding to the threshold voltage for LC domain reorientation. At this voltage, the electric force from the external field becomes equal to the anchoring force holding the LC to the initial orientation ($\varphi_i = 0$), and the directors start jumping to the next preferential orientation ($\varphi_f = 60$ °). Considering the distance between electrodes ~ 10 µm, the threshold electric field intensity $E_{th} = 1.2$ V/µm was estimated, which is much lower than that of bistable LC on four-fold symmetry patterned polyimide reported by Kim, et al.[30] This measurement was repeated to determine the threshold voltage on another sample with different shape of the grain, and similar threshold field was obtained. (Figure S21-22) Further increasing the field strength converted grain 4 to entirely green. The difference of the threshold voltage can be attributed to the defects of the grain. In figure 5(b), changing the direction of the field from right to left, contrary to above condition, all three grains converted to green including grain 4 at ΔV = 12 V. Further increase in the field strength converts all three grains to orange. In order to explain this behavior, we suggest a model with a tilted crystal orientation of h-BN as shown in in Figure 5(c). Three armchair directions were indicated as 'a', 'b', and 'c'. Assuming that initial orientation of LC was 'a', LC is rotated to 'b' clockwise by the field to +*x* direction. When the field is applied to -*x* direction, LC is rotated counterclockwise to 'c' at low field, but eventually moves to 'b' at high field. This demonstration confirms that all three states can be achieved by controlling the field strength and direction with two in-plane electrodes. On the other hand, no change on grain 2 can be explained by the fact that the LCs on the grain were initially oriented parallel to the external field.

The azimuthal anchoring strength of the liquid crystal/h-BN interface can be estimated from the threshold voltage. As the grain size of the h-BN is large, we can assume that the anchoring extrapolation length is smaller than the LC domain size.[30] The helical structure of the LC layers can be described by the azimuthal angle of the director, φ, which is a function of the distance from the surface. The anchoring energy is mainly determined by the azimuthal angle difference $\Delta\varphi$ at the interface. The anchoring energy is expressed by the Rapini-Papoular function[31, 32] $W(\Delta\varphi) = (W_o/2)\sin^2(\Delta\varphi)$ in general, where $W_o$ is the anchoring strength. In this hexagonal symmetric substrate, the anchoring energy will have a maximum value at $\Delta\varphi = \pi/6$. As the electric force reaches the threshold point, the LC director jumps to a next preferential orientation, resulting in color switching.

The threshold field for bistable LC molecules on a fourfold symmetrical surface was calculated by Kim et al. using the minimization of free energy with Rapini-Papoular anchoring energy[30] as

$$E_{th} = W_0/(2\sqrt{K_{22}\Delta\varepsilon}), \qquad \ldots \quad (7)$$

Where $K_{22}$ is the twist constant ($2 \times 10^{-12} N$) and $\Delta\varepsilon$ is the dielectric anisotropy ($7 \times 10^{-11}$ F/m)[33]. Resultant anchoring strength, $W_0$, was estimated to be $\sim 10^{-4}$ J/m$^2$. The anchoring strength with the hexagonal symmetry ($\Delta\varphi = \pi/6$) is approximately estimated as $W_o \approx 2.8 \times 10^{-5}$ J/m$^2$ from Eq. (7) and the measured threshold electric field $E_{th} = 1.2$ V/μm, which is stronger than that of the rubbed PMMA layer, but weaker than that of a nanoscale patterned polyimide layer.[34] This result shows that the anchoring strength was enough to be utilized for display applications. It is notable that this anchoring force is not only based on the topographical structure of the surface, but also based on the crystallographic structure of the hexagonal lattice.

Using DFT calculations, we checked external field effect on the orientation of a 5CB molecule on an h-BN nanoribbon (width ~2 nm) in the zigzag and the armchair direction (Figure S10). The two optimized structures were obtained in the absence of external electric field, and then a uniform external electric field was applied to the structure. As described earlier, the adsorption energy of the 5CB molecule along the armchair direction is stronger than that in the zig-zag direction in the absence of applied electric field. In contrast, the orientation of the 5CB molecule prefers to the zig-zag direction, when the intensity of the electric field is larger than 0.03 V/Å. When placed in an external electric field, an object with an electric dipole moment tends to align parallel to the field to make potential energy lower. As this simulation is based on single LC molecule and nanosized substrate, the threshold electric field was estimated to be much stronger than the experimental result with collective LC dipole array.

## Conclusion:

In this work, an innovative method for realizing multistable states of LC molecules on grains of h-BN has been proposed. The alignment of the LC was tuned via planar electric field. Several states with a variety of colors were confirmed by just varying the intensity of DC pulses. From electro-optical measurements, possibilities of novel display of various colors on a single pixel have been suggested. The new design paradigm presented in this work, provides a novel substitute approach towards the realization of filter-free non-volatile displays without backlight unit. Owing to the molecular alignment symmetry of the LC on the hexagonal lattice, we insist that an ultra-high resolution display with a molecular-scale pixel can be realized.

**Appendix A:**

From

$$\psi = \frac{a - ic}{a + ib} e^{2i\theta}$$

Considering $r = \sqrt{a^2 + b^2}$, $r' = \sqrt{a^2 + c^2}$, $g = \tan^{-1}\frac{b}{a}$ and $g' = \tan^{-1}\frac{-c}{a}$, we obtain

$$\psi = \frac{r'}{r} e^{i(g' - g + 2\theta)}$$

As the analyzer was set in y-axis, imaginary part of $\psi$ should be proportional to the detected intensity.

$$\text{Im}(\psi) = \frac{r'}{r} \sin(g' - g + 2\theta).$$

Or,

$$\text{Im}(\psi) = \frac{\sqrt{a^2 + c^2}}{\sqrt{a^2 + b^2}} \sin(\tan^{-1}\frac{-c}{a} - \tan^{-1}\frac{b}{a} + 2\theta).$$

## Methods:

### Synthesis of Boron Nitride (BN), Graphene and MoS$_2$:

Hexagonal BN films were grown on Cu foil (Alfa Aesar, 99.8% pure, 25μm thick) using a conventional thermal CVD mechanism. (Figure S1) Mechanical and electrical polishing was applied to the Cu foil, and it was annealed at 990 °C for 30 min with H$_2$ gas with 5 SCCM. After cleaning, h-BN was synthesized with borazine gas and hydrogen at 997 °C for 30 min. The furnace was cooled from 997 to 500 °C at a rate of ~35 °C/min. Ammonia borane (Sigma-Aldrich, 97% pure) was used as a precursor to form the h-BN film. The ammonia borane was thermally decomposed into hydrogen, aminoborane, and borazine at a temperature range from 80 to 120 °C. The h-BN films were transferred to secondary substrates, using a standard wet transfer method using poly methyl methacrylate (PMMA). (Figure S3) Graphene was grown using a conventional

CVD process with hydrogen and methane as carrier and precursor, respectively. $MoS_2$ was grown using a two-step process- first, molybdenum was deposited using sputtering and second, CVD was employed in a sulfur environment to synthesize $MoS_2$ films.

**LC Alignment:**

Commercially available nematic liquid crystal (5CB-Sigma Aldrich) was used. A 0.5µl liquid crystal was spin coated (3000 rpm) for 60 s on CVD-grown boron nitride and graphene films in order to make the LC wet the surface. The liquid crystal molecules were directly aligned along the surface of the h-BN and domain orientation/boundaries can be observed under POM with cross polarizers. The sample was heated in order to reach thermal equilibrium of the LC anchoring interaction. First, it was observed at room temperature, and then the temperature was raised to 60 ºC (higher than clearing point) for 15 minutes followed by cooling at ambient temperature. An LC cell was fabricated with a rubbed layer on the top electrode and polyvinyl alcohol (PVA) was spin coated on a glass slide. The film was uni-directionally rubbed and was placed over the LC-coated surface in order to confine the anchoring orientation at the top layer. This gap was controlled by polystyrene beads to obtain a 1–2 µm thick LC cell.

**Characterization:**

The LC alignment, boundaries and domain orientation were observed using a polarizer optical microscope (BX-51, Olympus Corp.). The axis of the polarizer was adjusted to be perpendicular to the analyzer. Raman spectroscopy with an excitation wavelength of 514 nm was used and power was kept below 1.0 mW to avoid laser-induced heating. The laser spot size of the Raman spectroscope was $1 \pm 0.2$ µm. The AFM images were obtained in non-contact mode in ambient conditions using a commercial AFM (n-Tracer, NanoFocus Inc.).

**Electro-optical measurements:**

Chrome masks were prepared with different types of electrodes. Optical lithography with a lift-off process was employed using a positive photoresist to pattern the electrodes on the silicon wafers. To investigate this electrical pulse response of aligned LC molecules, a digital signal generator was equipped with a digital-to-analog data acquisition interface card (National Instrument's PCI Express 6353). National Instruments' Data Measurement Services Software (NI-DAQmx) with LabVIEW was used to program pulse voltage sequences with a period of 2.5 ms. The voltage source unit and optical microscope image were synchronized by a PC program to allow correlated image analysis.

**Computational Details:**

Using DFT, we performed first-principles calculations to determine the favorable adsorption configurations of a 4-cyano-4'-pentylbiphenyl ($C_{18}H_{19}N$, also called 5CB) molecule and to calculate the atomic and electronic properties of them. We used the projector-augmented wave pseudopotential implemented in Vienna ab-initio simulation package (VASP) code[21, 22] and a plane-wave basis set with the cut-off energy of 500 eV. For exchange-correlation functional, the generalized gradient approximation (GGA) proposed by Perdew, Burke, and Ernzerhof[23] was employed. We also included the DFT-D2 method[24] for the van der Waals correction. The $6 \times 8$ h-BN supercell of an orthorhombic primitive cell was adopted as a substrate for liquid crystal molecules. In the primitive cell of h-BN, its optimized lattice parameters, *a* and *b,* were 4.35 Å and 2.51 Å, respectively. The lattice parameter, c, was chosen to be 24 Å. For the $6 \times 8$ supercell, the optimized lattice parameters, *a* and *b,* were 26.12 Å and 20.11 Å, respectively. Only the Γ-point was used in the $6 \times 8$ supercell calculation. AL model systems were relaxed until residual forces acting on each atom were smaller than 0.02 eV/Å.

**ACKNOWLEDGMENT**

This research was supported by the Priority Research Centers Program (2010-0020207) and (2017R1A2B4002379) through the National Research Foundation of Korea (NRF) funded by the Ministry of Education. This work was also supported by the Human Resources Development of the Korea Institute of Energy Technology Evaluation and Planning (KETEP) grant funded by the Korea government Ministry of Trade, industry & Energy (No. 20164030201340).

**Author contributions:** Y.S. conceived and designed the experiments. M.A.S. and Y.S. performed the device fabrication, characterized the device, and wrote the paper. J.L. and G.K. carried out the computer simulation and theoretical calculations. I.A, S.H.P., M.F.K., S.H, J.J, J.E., and C.H., prepared samples and analyzed the data. All authors discussed the results and commented on the manuscript.

**Competing interests:** All authors declare that they have no competing interests.

**Data Availability statement:** All data needed to evaluate the conclusions in the paper are present in the paper and/or the Supporting Information. Additional data related to this paper may be requested from the authors.

# Supporting Information

Supporting Information[35, 36] with supplementary data set is available for this article.

Supporting video shows a real-time POM image of LC device with three electrode.

# One-sentence summary

Liquid crystal alignment on 2D materials could be utilized to fabricate nonvolatile and high definition displays.